# Spatio-temporal Trajectory Dataset Privacy Based on Network Traffic Control

Qilong Han, Qianqian Chen, Kejia Zhang, Xiaojiang Du and Nadra Guizani

**Abstract**—Collection of user's location and trajectory information that contains rich personal privacy in mobile social networks has become easier for attackers. Network traffic control is an important network system which can solve some security and privacy problems. In this paper, we consider a network traffic control system as a trusted third party and use differential privacy for protecting more personal trajectory data. We studied the influence of the high dimensionality and sparsity of trajectory data sets based on the availability of the published results. Based on similarity point aggregation reconstruction ideas and a prefix tree model, we proposed a hybrid publishing method of differential privacy spatiotemporal trajectory data sets APTB.

**Index Terms**—network traffic control, mobile social network, differential privacy

## 1 INTRODUCTION

TRAFFIC control, especially network traffic control, which plays a significant role in network systems, is able to solve some security and privacy problems such as internal information leakage as well as the personal information disclosure. The main purpose of network traffic control is to satisfy users' requirements of networking, to provide users with better quality of network services, and to protect the transmission safety of monitoring data at the same time. The network traffic control systems not only encrypt data transmission, but also intercept and block potential unsafe protocols. They also block illegal accesses as well as virus invasions.

Mobile social networks now are widely used for communication and interaction among people[5]. Seamless mobile computing, key management[14-16] and social computing together, greatly enhances the authenticity, regional and interactive users in a real-time mobile terminal and the user is bound to ensure the authenticity of the social network. In addition, the introduction of location information brings social networking service diversification and individuation that can provide real-time interaction between users. We witness new types of information that involve spatio-temporal trajectory data.

We can not deny the importance of spatio-temporal trajectory data. It can help research on user behavior theories and on issues such as network traffic burst prediction and network resource allocation. With the current increase of location service applications, however, the research on the privacy protection of location trajectory data has become a popular research topic and has produced a collection of results [1-3].

Differential privacy is widely studied[4-6] for the privacy and security issues of statistical data. For a trajectory data set, we will use differential privacy for manipulating the spatio-temporal trajectory data while monitoring traffic of the mobile terminal. This is to protect personal privacy as well as keep data availability. Applying the differential privacy model to the publication of trajectory data sets is of great significance to the privacy and security that's studied in [17-20] for mobile information.

Trajectory privacy not only includes the trajectory itself, but also includes sensitive location points in the trajectory and other personal information that can be mined through the trajectory. Trajectory data sets can be divided into spatial trajectory data sets and spatio-temporal trajectory data sets according to their type. Differential privacy algorithms[7,8] such as prefix

————————————————

- *Qilong Han is with Harbin Engineering University, Harbin, China E-mail:hanqilong@hrbeu.edu.cn*
- *Qianqian Chen is with Harbin Engineering University, Harbin, China E-mail:chenqianqian@hrbeu.edu.cn*
- *Kejia Zhang(corresponding author) are with Harbin Engineering University, Harbin, China E-mail:kejiazhang@hrbeu.edu.cn*
- *Xiaojiang Du：Dept. of Computer and Information Sciences, Temple University, Philadelphia PA 19122, USA, email: dxj@ieee.org*
- *Nadra Guizani, Dept. of Electrical and Computer Engineering, Purdue University, USA. E-mail: nguizani@purdue.edu*

and n-gram are published for spatial trajectory data sets, and differential privacy algorithms such as GST for spatio-temporal trajectory data sets. The biggest difference between the spatial trajectory data set and the spatio-temporal trajectory data set is whether the position points in the trajectory contain time attributes. Although there is only one time attribute in the definition of the spatio-temporal trajectory data set, it will have a significant impact on the efficiency of the publishing algorithm and the accuracy of the published data set.

In this paper, we propose a method of hybrid publishing of spatial-temporal differential privacy trajectory data sets, aiming at the shortcomings of existing differential privacy trajectory data sets.

## 2 PROBLEM SETUP

We use the prefix tree model to publish the spatio-temporal trajectory dataset that ultimately satisfies the differential privacy. The temporal attribute of the spatio-temporal location points leads to many nodes in the prefix tree and the problem of excessive noise added. The idea of partition and merge is used to propose a node aggregation reconfiguration. The noise prefix tree construction algorithm APTB, can effectively reduce the overall noise injection amount of the prefix tree under the premise of guaranteeing differential privacy.

### 2.1 Problem

Compared to ordinary spatial trajectory datasets, there are more spatial-temporal trajectory data points in the time-entry attribute. This forms a large number of branches in the prefix tree, resulting in excessive accumulated noise in the prefix tree. This also reduces the availability on releasing data set.

In the prefix tree of this paper, we consider the root node as a virtual node. The root is located at the $0^{th}$ level of the tree. The branch from the root node to the current node represents a prefix, and the branch from the root node to the leaf node is called a path. Each node in the tree consists of triple($L,\varepsilon,c$), where L represents the spatio-temporal location represented by the node, $\varepsilon$ represents the privacy budget assigned to the node, and c represents the number of tracks in the data set that contain the node prefix. For example, TABLE 1 shows a trajectory dataset aggregated over time and space. Its real prefix tree is shown in Fig1.

TABLE 1
SPATIO-TEMPORAL TRAJECTORY DATASETS AGGREGATED OVER TIME AND SPACE

| No. | trajectory |
|---|---|
| 1 | $L_1 \rightarrow L_2$ |
| 2 | $L_1 \rightarrow L_3 \rightarrow L_4$ |
| 3 | $L_2 \rightarrow L_3$ |
| 4 | $L_3 \rightarrow L_5$ |
| 5 | $L_2 \rightarrow L_5 \rightarrow L_6$ |
| 6 | $L_3 \rightarrow L_4 \rightarrow L_5$ |
| 7 | $L_3 \rightarrow L_5 \rightarrow L_6$ |
| 8 | $L_4 \rightarrow L_5$ |
| 9 | $L_4 \rightarrow L_5 \rightarrow L_6$ |
| 10 | $L_4 \rightarrow L_5$ |
| 11 | $L_5 \rightarrow L_6$ |
| 12 | $L_5 \rightarrow L_6$ |
| 13 | $L_4 \rightarrow L_7$ |
| 14 | $L_5 \rightarrow L_6$ |
| 15 | $L_5 \rightarrow L_7$ |

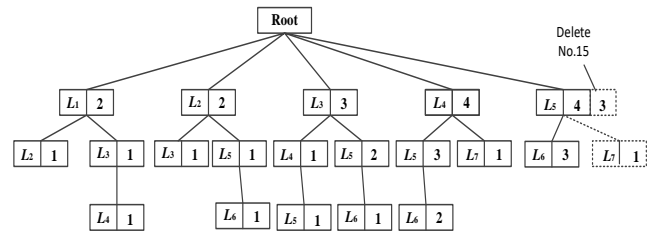

Fig.1 Real prefix map generated from TABLE 1

Adjacent data sets refer to the spatio-temporal trajectory data set recorded only by one trajectory, so the global sensitivity of the query is 1. Therefore, we can use Laplacian mechanism to add noise for implementing differential privacy protection for trajectory datasets. The amount of noise added for each node is subject to the Lap(1/ε) distribution. However, the spatio-temporal trajectory dataset has high dimensional characteristics. If noise is directly added, each layer of the prefix tree accumulates a large amount of noise data, resulting in a reduced availability of the final published track data set.

A feasible solution to solve the problem mentioned previously is to combine nodes with similar count values into a coarse node for nodes assigned to the same privacy budget in the same layer. While the coarse node count becomes the sum of the counts of the member nodes, it is only adding a separate Laplacian noise to the coarse node that can guarantee the differential privacy of all the member nodes. At the time of publication, the count of member nodes is the average number of coarse node noise counts. Assuming that the coarse node contains m member nodes, we can know that the average amount of noise added by each member node becomes 1/m before the merge. This solution can reduce the amount of noise added effectively and respond to longer-range queries.

However, since the counts of the member nodes at the time of publication use the average number, not only noise errors but also the reconstruction error caused by the average affects the accuracy of the member node count. Therefore, in order to truly improve the accuracy of the published data set, the relationship between noise error and reconstruction error must be balanced.

## 2.2 APTB Overview

To solve this problem, we propose a noise prefix tree construction algorithm APTB[10] based on node aggregation reconstruction.

The main idea of the APTB is to reduce cumulative error and transfer error first by limiting the height of the prefix tree and the number of nodes that can be extended downward in each layer. Secondly, for nodes assigned to the same privacy budget in the same layer of the prefix tree, they are first aggregated according to the count value, and then use the exponential mechanism to select neighboring nodes with similar count values for reconstructing and re-assembling, and adding noise to the reconstructed coarse nodes. Taking the average number of coarse node noise counts as the count of member nodes. Finally, we process the constructed noise prefix tree in a consistent manner, and generate a final released spatio-temporal trajectory data set based on the processed results.

When the noise prefix tree is expanded downwards, if the noise count of a node is very small, according to the nature of the Laplacian mechanism, its true count value will also be small. Continuing to expand this type of node will make the number of nodes in the tree grow exponentially, which not only reduces the efficiency of the algorithm and increases the calculation cost, but also increases the transmission error and accumulated noise amount of the prefix tree. Therefore, we limit such nodes by thresholds θ. For noise-counted nodes which c(Ñ)<θ, we mark them as leaf nodes and do not continue to expand to the next level.

Assuming h=3, θ=2, the noise prefix tree constructed from the spatio-temporal trajectory dataset shown in Table -1 is shown in Fig.2. Only some of the nodes are shown and the construction of the unshown part is similar to that of other nodes.

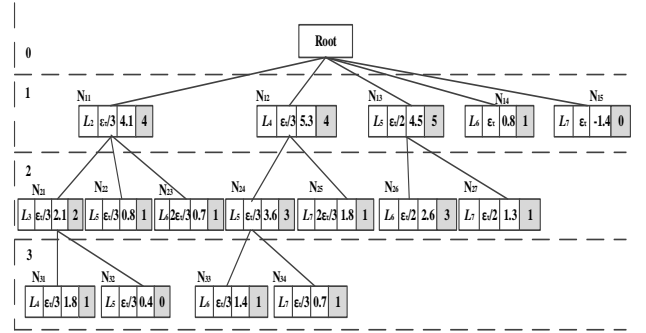

Fig.2 Noise Prefix Tree

The noise prefix tree construction algorithm APTB and subsequent consistency processing will be described in detail below.

## 3 NODE PRIVACY BUDGET ALLOCATION

Before constructing the noise prefix tree, we calculate the height *h* of the tree first, where h does not contain the root node, Root. The larger the *h* is, the longer the maximum length of the final released dataset trajectory. Also, the included sequence information becomes richer, but it increases the risk of privacy leakage. Moreover, the value of h will also affect the total amount of noise in the prefix tree. The greater the value of h, the greater the overall noise level. However, if the value of h is too small, the published track data set will lose part of the sequence information, making it less capable of responding to data mining tasks. Therefore, a reasonable h must be set to balance the privacy and availability of the published data set. However, most of the publishing algorithms require the publisher to set it up based on the actual characteristics of the data set.

We then divide the privacy budget into two equal parts, one for adding noise to the maximum length of the trajectory, and the other for adding noise to the count of the trajectory length. In the second line of the algorithm, the reason why the noise is added to the maximum track length l is to consider such a case: Assuming that in $\bar{D}$ the number of tracks with a length of l is 1, we use tr to represent this track, if we delete tr to form a contiguous dataset $\bar{D}'$ by $\bar{D}$, then the structure of the trace length count value vector in $\bar{D}'$ will change, thus destroying differential privacy.

In order to better utilize the privacy budget and reduce waste, we have proposed a new privacy budget allocation strategy based on the characteristics of the spatio-temporal trajectory dataset. For a node N to be allocated a private budget, we use Tree(N) to represent a subtree with N as the root node. When assigning the privacy budget, we calculate theoretical maximum expansion layer number maxExpand(N) of N. The greater of maxExpand(N), means the higher the theoretical height of Tree(N), and the smaller the assigned privacy budget value.The smaller of maxExpand(N), the greater the allocated privacy budget. For the calculation of maxExpand(N), the set of candidate child nodes sonSet(N) of N is first required.

After computing the candidate set of child nodes sonSet(N), if the result is sonSet(N) = null, then maxExpand(N) = 0; otherwise, select the child node s with the largest number of expandable layers from sonSet(N),

Since the value of Tree(N) depends not only on maxExpand(N), but also by the noise prefix tree height h and the number of layers in which N is located. Although the distribution of the above privacy budgets cannot completely avoid waste due to the impact of noise thresholds θ, it still plays a significant positive role in improving the availability of published data sets. In future work, we will further optimize the distribution of privacy budgets based on the effect of θ on the height of the prefix tree.

## 4  NODE AGGREGATION

In order to solve the problem that the high dimensionality of spatio-temporal trajectory dataset leads to adding too much noise data, we propose the idea of partitioning and merging nodes to assign to the same privacy budget in the same layer of the noise prefix tree and reduce noise errors. The pre-reconstruction plus noise technique based on partitioning and merging has been widely used in the field of differential privacy histogram publications and its validity and privacy have been proved theoretically. These theories are also applicable to the construction of noise prefix tree. Moreover, the nodes in the noise prefix tree have a great similarity with the histogram. We can map the prefixes represented by the nodes to the attributes of the histograms and map the counts of the nodes to the bucket frequency of the histogram. When the pre-reconstruction technique is used to process the nodes in the same layer of the noise prefix tree, the outliers in the node count value will also bring a large reconstruction error to the publishing result, and this error will be passed down [11-13]. In the deeper construction of the prefix tree, the error becomes larger and larger. Therefore, compared to the impact of outliers on the accuracy of the published data of the differential privacy histogram, its impact on the construction of the prefix tree is much greater.

For eliminating this effect, we have to aggregate the node with similar counts into the same cluster, and aggregate the outliers into separate clusters. When the nodes are combined and reconstructed, clusters are used as the unit, only the similar nodes in the same cluster are merged, thereby eliminating the influence of outliers on node reconstruction.

We should also compare the true counter value plus the noise size when the nodes are aggregating. This process needs to input a count value interval distance δ, which can be also called outlier distance. Nodes whose noise count value is larger than the outlier distance will not be assigned to the same one.

## 5 NODE RECONSTRUCTION

After the aggregating and sorting nodes according to the noise count value of the node, the cluster-units are also used to combine and reconstruct neighboring nodes with similar count values in the cluster to reduce the amount of noise added. We considered combining nodes with similar count values in the prefix tree and then adding noise can effectively reduce the total amount of noise added. However, the merging will bring about new reconstruction errors. In order to effectively balance the noise error and the reconstruction errors, the merging operation is more reasonable.

In this part, before adding noise to the coarse nodes, we should begin with selecting the neighboring nodes with similar count values to perform a merge operation, and then increase the number of merged nodes through multiple cycles. When choosing the merging neighbors, in order to ensure differential privacy, we use the exponential mechanism to choose and use the similar distance of nodes as the scoring function of the index the selection of the node's consolidation scheme, the selection operation loop is performed k times, where k is the number of nodes in the cluster. The smaller the similarity distance between two adjacent nodes, the higher the score and the greater the probability of being selected for merging.

When the number of nodes C, which represents the node cluster collection that assigned to the same privacy budget in layer i, is less than or equal to 2, the output scheme selected by the exponential mechanism is one and only one, so no choice of exponential mechanism is required. Therefore, when k ≤ 2, this step does not consume the privacy budget. The fifteenth line of the algorithm is the process of adding noise to the merged node, where in $\tilde{N}_j^i$, the count value is the noise result $\tilde{n}_j^i$, but the other node information is the same to $\tilde{n}_j^i$. When noise is added, only noise is added to the coarse node.

## 6 NOISE THRESHOLD AND CONSISTENCY PROCESSING

In the noise prefix tree, the error of the parent node is passed down to the child node, causing an accumulation of errors. The biggest source of this kind of error is a node with a true counter value of 0 in itself. Since the noise is added and its count value is greater than the threshold θ, it can continue to expand downwards. Therefore, we need to design a reasonable threshold to limit the generation of such nodes to the maximum extent without affecting the expansion of nodes with larger true counts. In terms of choice, literature [9] gives an effective solution, and we make some improvements based on it, making the algorithm work better in the context of this article. However, unlike [9], we do not set the same threshold for all nodes in the tree. Instead, we update the value according to the privacy budget assigned by the node and the aggregation reconfiguration of the cluster.

When adding noise to the node's count value, we suppose that there are k nodes in C, and the probability that a node with a true counter value of 0 exceeds after adding noise is Pθ, then, we hope $kP_\theta < 1$ for each C. Through experiments on real data sets, it is found that the threshold calculation method described above can effectively control the number of nodes whose real count value in each layer of the noise prefix tree continues to expand downward.

Before generating a spatio-temporal trajectory dataset that satisfies differential privacy based on a noise prefix tree, we must adjust the node count value in the noise prefix tree to meet the consistency requirement. Because the size of the Laplacian noise added by the tree node is only related to the privacy budget when the global sensitivity is determined. If the child node is assigned a small private budget, it is likely that the Laplacian noise added is too large to exceed its parent node's count. In order to a better solution of this problem, we give the concept of node consistency, as defined in Definition 2.2.

**Definition 2.2**. *Given a noise prefix tree node N and its child node set S(N), the count value of N is set to n. Then the node's consistency is defined as follows:*

*1) For any $s \in S(N)$, its count value $s \leq n$;*

*2) $n \geq \sum_{s \in S(N)} s$;*

*3) n is an integer greater than or equal to 0.*

In order to satisfy the definition for all nodes in the noise prefix tree, we make the following adjustments to the count value of the node:

1) Starting from the first layer of the prefix tree layer by layer, for each node N of the i-th layer, if its count value n is less than 0, then let n=0; if there are satisfy $\sum s = n$, count value $s_i$ in each child node $s_i$ in N will be adjusted.

2) Round off the counts of all nodes in the noise prefix tree.

In Fig.2, the count value of each node is shown as a shaded part after adjusting. it is necessary to generate a finally released space-time trajectory data set satisfying differential privacy according to the tree structure. The space-time release trajectory data set is shown in TABLE 2 (excluding the part with the prefix $L_1$ and $L_3$).

TABLE 2
PUBLISHED SPATIO-TEMPORAL TRAJECTORY DATASETS

| No. | trajectory |
|---|---|
| 1 | $L_2 \rightarrow L_3 \rightarrow L_4$ |
| 2 | $L_2 \rightarrow L_3$ |
| 3 | $L_2 \rightarrow L_3$ |
| 4 | $L_2 \rightarrow L_5$ |
| 5 | $L_2 \rightarrow L_6$ |
| 6 | $L_6$ |
| 7 | $L_4 \rightarrow L_5 \rightarrow L_6$ |
| 8 | $L_4 \rightarrow L_5 \rightarrow L_7$ |
| 9 | $L_4 \rightarrow L_5$ |
| 10 | $L_4 \rightarrow L_5$ |
| 11 | $L_4 \rightarrow L_7$ |
| 12 | $L_5 \rightarrow L_7$ |
| 13 | $L_5 \rightarrow L_6$ |
| 14 | $L_5 \rightarrow L_6$ |
| 15 | $L_5 \rightarrow L_6$ |
| 16 | $L_5$ |

## 7 CONCLUSION

With the rapid development of mobile social networks, it has become easier to collect user's location and trajectory information. Through the excavation of trajectory data, a large number of potential models can be discovered. However, the trajectory contains rich personal privacy. The trajectory data can be used to obtain private information such as the user's lifestyle and home address. As people's privacy awareness gradually improves, privacy protection for the release of trajectory data sets cannot be delayed. Network traffic control is an important network systems which can solve some security and privacy problems. As a new privacy standard, differential privacy has gradually been applied to various fields of privacy protection due to its theoretical solidity. In this paper, we consider the network traffic control system as trusted third party and use differential privacy in it for protecting personal trajectory data more. However, the existing differential privacy trajectory data set publication method mainly protects the data set with a small location points and does not consider the time attribute of the location points. In the publication of spatio-temporal trajectory data sets with large time-point attributes and location points.

The research content of this paper still has certain optimization space. In the next step, we will start with the following and continue to conduct more in-depth research: In the privacy budget allocation of the noise prefix tree node, the influence of the noise threshold θ on the height of the tree is not considered. Next, it is necessary to consider how to add the θ parameter to the privacy budget allocation algorithm to further reduce the waste of the privacy budget.

**REFERENCES**

[1] Wang E K, Ye Y. A New Privacy-Preserving Scheme for Continuous Query in Location-Based Social Networking Services. International Journal of Distributed Sensor Networks, 2014(4):1-11P

[2] Gao S, Ma J, Shi W, et al. LTPPM: a location and trajectory privacy protection mechanism in participatory sensing. Wireless Communications & Mobile Computing, 2015, 15(1):155–169P

[3] Abul O, Bonchi F, Nanni M. Never Walk Alone: Uncertainty for Anonymity in Moving Objects Databases. International Conference on Data Engineering. IEEE Computer Society, 2008:376-385P

[4] Andrienko G, Andrienko N, Giannotti F, et al. Movement data anonymity through generalization. Transactions on Data Privacy, 2010, 4(2):51-60P

[5] Chen R, Mohammed N, Fung B C M, et al. Publishing set-valued data via differential privacy. Proceedings of Very Large Databases(VLDB). Seattle, 2011:1087- 1098P

[6] Chen R, Fung B C M, Desai B C. Differentially Private Trajectory Data Publication. Computer Science, 2011

[7] Dwork C, McSherry F, Nissim K, et al. Calibrating noise to sensitivity in private data analysis. Proceedings of the 3rd Theory of Cryptography Conference (TCC). New York, 2006:363-385P

[8] McSherry F, Talwar K. Mechanism design via differential privacy. Proceedings of the 48th Annual IEEE Symposium on Foundations of Computer Science (FOCS). Providence, 2007:94-103P

[9] Chen R, Acs G, Castelluccia C. Differentially Private Sequential Data Publication via Variable-Length N-Grams. ACM Computer and Communication Security (CCS), 2012:638-649P

[10] Shao D, Jiang K, Kister T, et al. Publishing Trajectory with Differential Privacy: A Priori vs. A Posteriori Sampling Mechanisms Database and Expert Systems Applications. Springer Berlin Heidelberg, 2013:357-365P

[11] He X, Cormode G, Machanavajjhala A, et al. DPT: differentially private trajectory synthesis using hierarchical reference systems. Proceedings of the VLDB Endowment, 2015, 8(11):1154-1165P

[12] Hua J, Gao Y, Zhong S. Differentially private publication of general time-serial trajectory data. Computer Communications. IEEE, 2015:549-557P

[13] Han Q, Shao B, Li L, et al. Publishing histograms with outliers under data differential privacy. Security & Communication Networks,2016,9(14): 2313-2322P

[14] Du X, Guizani M, Xiao Y and Chen H. H, "A Routing-Driven Elliptic Curve Cryptography based Key Management Scheme for Heterogeneous Sensor Networks," IEEE Transactions on Wireless Communications, Vol. 8, No. 3, pp. 1223 - 1229, March 2009.

[15] Xiao Y, Rayi V, Sun B, Du X, Hu F, and Galloway M, "A Survey of Key Management Schemes in Wireless Sensor Networks," Journal of Computer Communications, Vol. 30, Issue 11-12, pp. 2314-2341, Sept. 2007.

[16] Du X, Xiao Y, Guizani M, and Chen H. H, "An Effective Key Management Scheme for Heterogeneous Sensor Networks," Ad Hoc Networks, Elsevier, Vol. 5, Issue 1, pp 24–34, Jan. 2007.

[17] Wu L, Du X, and Fu X, "Security Threats to

Mobile Multimedia Applications: Camera-based Attacks on Mobile Phones," IEEE Communications Magazine, Vol. 52, Issue 3, March 2014.

[18] Wu L, and Du X, "MobiFish: A Lightweight Anti-Phishing Scheme for Mobile Phones," in Proc. of the 23rd International Conference on Computer Communications and Networks (ICCCN), Shanghai, China, August 2014.

[19] Huang X, Du X, "Achieving big data privacy via hybrid cloud," in Proc. of 2014 IEEE INFOCOM Workshops, Pages: 512 – 517

[20] Du X and Chen H. H, "Security in Wireless Sensor Networks," IEEE Wireless Communications Magazine, Vol. 15, Issue 4, pp. 60-66, Aug. 2008.